\documentclass[aps,prd,twocolumn,showpacs,amsmath,amssymb]{revtex4-1}
\usepackage{amsmath} \usepackage{graphicx} \usepackage{subfigure}
\usepackage{epstopdf} \usepackage{color} \usepackage{multirow}
\usepackage{setspace} \usepackage{overpic} \usepackage{amssymb}
\usepackage{ulem}
\usepackage{siunitx}

\usepackage[bookmarksnumbered, pdfstartview=FitH,colorlinks,urlcolor=blue, citecolor=blue,linkcolor=blue] {hyperref}
\usepackage{lineno}
\usepackage{bm}
\usepackage{rotating}
\usepackage{xcolor}%0824
\usepackage{makecell}
\usepackage{mathtext}
\usepackage{mathrsfs}
\usepackage[utf8]{inputenc}
\usepackage{threeparttable}
\let\oldequation\equation
\let\oldendequation\endequation
\renewenvironment{equation}
 {\linenomathNonumbers\oldequation}
 {\oldendequation\endlinenomath}

\include{def-com-new}
\begin{document}
%\linenumbers

\title{\boldmath Search for $h_c \to \pi^+\pi^-J/\psi$ via $\psi(3686)\to \pi^0h_c$ }

%% Saved at => 2024-07-05
\author{%%
\begin{small}
\begin{center}
M.~Ablikim$^{1}$, M.~N.~Achasov$^{4,c}$, P.~Adlarson$^{76}$, O.~Afedulidis$^{3}$, X.~C.~Ai$^{81}$, R.~Aliberti$^{35}$, A.~Amoroso$^{75A,75C}$, Y.~Bai$^{57}$, O.~Bakina$^{36}$, I.~Balossino$^{29A}$
% ...
\\
\vspace{0.2cm}
(BESIII Collaboration)\\
\vspace{0.2cm} {\it
$^{1}$ Institute of High Energy Physics, Beijing 100049, People's Republic of China\\
% ...
$^{3}$ Bochum  Ruhr-University, D-44780 Bochum, Germany\\
$^{4}$ Budker Institute of Nuclear Physics SB RAS (BINP), Novosibirsk 630090, Russia\\
% ...
$^{29}$ INFN Sezione di Ferrara, (A)INFN Sezione di Ferrara, I-44122, Ferrara, Italy; (B)University of Ferrara,  I-44122, Ferrara, Italy\\
% ...
$^{35}$ Johannes Gutenberg University of Mainz, Johann-Joachim-Becher-Weg 45, D-55099 Mainz, Germany\\
$^{36}$ Joint Institute for Nuclear Research, 141980 Dubna, Moscow region, Russia\\
% ...
$^{57}$ Southeast University, Nanjing 211100, People's Republic of China\\
% ...
$^{75}$ University of Turin and INFN, (A)University of Turin, I-10125, Turin, Italy; (B)University of Eastern Piedmont, I-15121, Alessandria, Italy; (C)INFN, I-10125, Turin, Italy\\
$^{76}$ Uppsala University, Box 516, SE-75120 Uppsala, Sweden\\
% ...
$^{81}$ Zhengzhou University, Zhengzhou 450001, People's Republic of China\\
\vspace{0.2cm}
% ...
$^{c}$ Also at the Novosibirsk State University, Novosibirsk, 630090, Russia\\
% ...
}\end{center}
\vspace{0.4cm}
\end{small}
}
%% ends here %%

%\input{authorlist_2023-11-13.tex}

\date{\today}

\def\TimesBF{\ensuremath{6.7\times 10^{-7}}}
\def\BF{\ensuremath{9.4 \times10^{-4}}}
\def\Totsys{\ensuremath{6.6}}

\begin{abstract}
 Using $(2712.4 \pm 14.3) \times 10^6~\psi$(3686) events collected with the BESIII detector operating at the BEPCII collider, we search for the hadronic transition $h_c \to \pi^+\pi^-J/\psi$ via $\psi(3686)\to \pi^0 h_c$. No significant signal is observed. We set the most stringent upper limits to date on the branching fractions $\mathcal{B}(\psi(3686)\to \pi^0 h_c)\times\mathcal{B}(h_c\to\pi^+\pi^-J/\psi)$ and  $\mathcal{B}(h_c \to \pi^+\pi^-J/\psi)$ at the 90$\%$ confidence level, which are determined to be \TimesBF and \BF, respectively.
\end{abstract}

\maketitle
	%%%%%%%%%%%%%%%%%%%%%%%%%%%%%%%%%%%%%%%%
	%          1. Introduction
	%%%%%%%%%%%%%%%%%%%%%%%%%%%%%%%%%%%%%%%%
\section{Introduction}
The study of charmonium decays is crucial to elucidate the
mechanism of quantum chromodynamics (QCD), as these states lie in the
transition region between non-perturbative and perturbative QCD.
Although QCD has successfully explained many aspects of strong
interactions, some known charmonium decay mechanisms remain
challenging, as documented in Ref.~\cite{theory_review_a}.

Following identification of spin-singlet $P$-wave charmonium
state $h_{c}(^1P_{1})$ in 2005~\cite{hc_a,hc_b}, extensive theoretical
and experimental efforts have been made to understand its
characteristics.  The study of $h_c$ remains difficult due to its
relatively low branching fraction (BF), because its production through
$1^{--}$ charmonia is suppressed~\cite{pdg2022}.

The first evidence for the $h_c$ state was reported by E835 at
Fermilab in the $p\bar{p} \to h_c \to \gamma \eta_c$
process~\cite{PRD032001}. Subsequently, the CLEO experiment presented
the first observation of the $h_c$ in a study of the cascade decay
$\psi(3686) \to \pi^0h_c, h_c \to \gamma \eta_c$~\cite{PRL102003},
measured its mass~\cite{PRL182003}, and provided evidence for its
multi-pion decay modes~\cite{PRD051106}.  The BESIII collaboration
made the first measurement of the absolute BFs,
$\mathcal{B}(\psi(3686) \to \pi^0h_c)$ and $\mathcal{B}(h_c \to \gamma
\eta_c)$~\cite{PRL132002}, which were later confirmed by
CLEO~\cite{PRD032008}.  With a larger $\psi(3686)$ data sample, recent
experimental measurements give $\mathcal{B}(\psi(3686) \to \pi^0h_c) =
(7.32\pm0.34\pm0.41)\times10^{-4}$ and $\mathcal{B}(h_c \to \gamma
\eta_c)=(57.66^{+3.62}_{-3.50}\pm0.58)\%$~\cite{PRD072007}.  These
indicate that the sum of the $h_c$ hadronic decay modes is
comparable to its radiative transition rate.

The hadronic transitions of $h_c$ provide a valuable opportunity to
investigate the spin-spin interaction between heavy
quarks~\cite{JPCS123}. The Feynman diagram of the hadronic transition
$h_c \to \pi^+ \pi^- J/\psi$ is depicted in Fig.~\ref{fig:fey}.
According to theoretical predictions based on a QCD multipole expansion,
the BF of $h_c \to \pi\pi J/\psi$ (including charged and neutral
modes) is predicted to be 2\%~\cite{PRD1210}. However, when
non-locality in time is neglected, the predicted BF decreases
significantly to 0.05\%~\cite{PRD1710}. Additionally,
Ref.~\cite{PRD074033} suggests that the BF for $h_c \to
\pi^+\pi^-J/\psi$ is approximately 40 times smaller than that for $h_c
\to \pi^0J/\psi$.

In 2018, the BESIII collaboration searched for the hadronic transition
of $h_c \to \pi^+\pi^-J/\psi$ using a dataset of $( 448.1 \pm 0.8)$
million $\psi(3686)$ events~\cite{CPC023001}. The upper limit on the
product of BFs, $\mathcal{B}(\psi(3686)\to \pi^0
h_c)\times\mathcal{B}(h_c\to\pi^+\pi^-J/\psi)$, was determined to be
$2.0\times 10^{-6}$ at the 90$\%$ confidence
level~(C.L.)~\cite{PRD052008}, which is in better agreement with the
theoretical prediction presented in Ref.~\cite{PRD1710}.  Four years
later, the BESIII Collaboration reported the upper limit on
$\mathcal{B}(h_c \to \pi^0J/\psi)$ which was determined to be $4.7
\times 10^{-4}$ at the 90\% C.L., utilizing 11 fb$^{-1}$ of $e^+e^-$
collision data taken at center-of-mass energies between 4.189 and
4.437 GeV~\cite{JHEP003}. According to Ref.~\cite{PRD074033}, this
result suggests that $\mathcal{B}(h_c\to\pi^+\pi^-J/\psi)$
could be smaller than $1.2 \times 10^{-5}$.

In this paper, based on $(2712.4 \pm 14.3) \times 10^6 ~\psi(3686)$ events~\cite{CPC023001, psi2S2021}, we present an updated analysis of a search for the hadronic transition $h_c \to \pi^+\pi^- J/\psi$.

 \begin{figure}[htbp]
 \label{fig:fey}
\begin{center}
\begin{minipage}[t]{1.0\linewidth}
\includegraphics[width=1\textwidth]{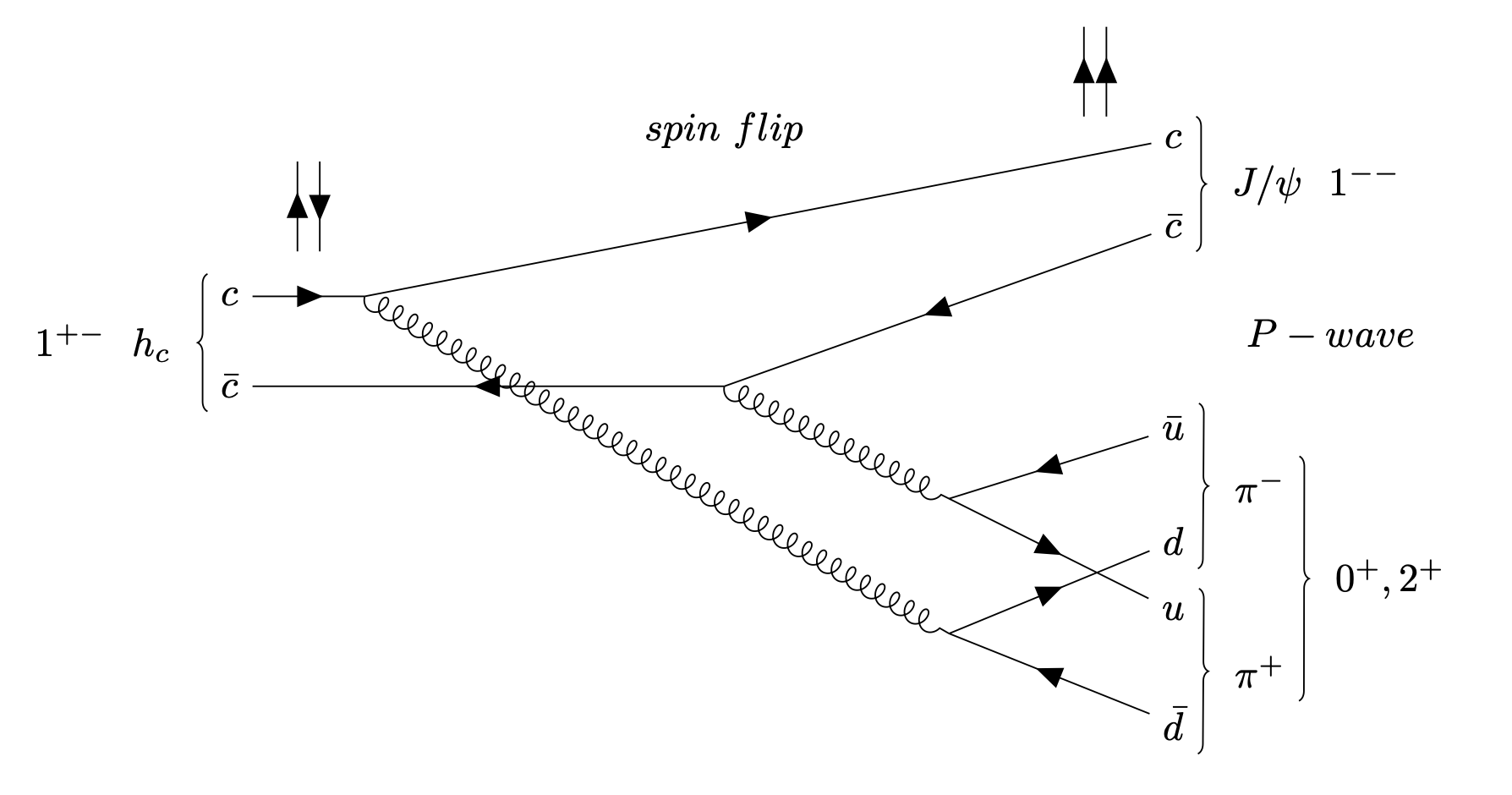}
\end{minipage}
\caption{The Feynman diagram for the spin flip hadronic transition $h_c \to \pi^+ \pi^- J/\psi$. }
 \label{fig:fey}
\end{center}
\end{figure}

	%%%%%%%%%%%%%%%%%%%%%%%%%%%%%%%%%%%%%%%%%%%%%%%%%%%%%%%%%%%%%%%%
	%          2. BESIII DETECTOR AND MONTE CARLO SIMULATION
	%%%%%%%%%%%%%%%%%%%%%%%%%%%%%%%%%%%%%%%%%%%%%%%%%%%%%%%%%%%%%%%%
\section{BESIII DETECTOR AND MONTE CARLO SIMULATION}
\label{sec:BES}

The BESIII detector~\cite{Ablikim:2009aa} records symmetric $e^+ e^-$
collisions provided by the BEPCII storage ring~\cite{CXYu_bes3} in the
center-of-mass energy range from 1.84 to 4.95~GeV, with a peak
luminosity of $1.1 \times 10^{33}\;\text{cm}^{-2}\text{s}^{-1}$
achieved at $\sqrt{s} = 3.773\;\text{GeV}$. BESIII has collected large
data samples in this energy
region~\cite{Ablikim:2019hff,EcmsMea,EventFilter}. The cylindrical
core of the BESIII detector covers 93\% of the full solid angle and
consists of a helium-based multilayer drift chamber (MDC), a
time-of-flight system (TOF), and a CsI(Tl) electromagnetic calorimeter
(EMC), which are all enclosed in a superconducting solenoidal magnet
providing a 1.0~T magnetic field. The solenoid is supported by an
octagonal flux-return yoke with modules of resistive plate muon
counters (MUC) interleaved with steel. The charged-particle momentum
resolution at 1~GeV$/c$ is 0.5\%, and the d$E/$d$x$ resolution is 6\%
for the electrons from Bhabha scattering. The EMC measures photon
energy with a resolution of 2.5\% (5\%) at 1~GeV in the barrel
(end-cap) region. The time resolution of the plastic scintillator TOF
barrel part is 68~ps, while that of the end-cap part is 110~ps. The
end-cap TOF system was upgraded in 2015 using multi-gap resistive
plate chamber technology, providing a time resolution of 60~ps, which
benefits $\sim83\%$ of the data used in this
analysis~\cite{tof_a,tof_b,tof_c}.

Monte Carlo (MC) simulated data samples produced with {\sc
  geant4}-based~\cite{geant4} software, which includes the geometric
description of the BESIII detector and the detector response, are used
to optimize the event selection criteria and estimate the signal
efficiency and level of background. The simulation models the
beam-energy spread and initial-state radiation in the $e^+e^-$
annihilation using the generator {\sc kkmc}~\cite{kkmc_a,kkmc_b}. The
inclusive MC sample includes the production of the $\psi(3686)$
resonance, the initial-state radiation production of the $J/\psi$
meson, and the continuum processes incorporated in {\sc kkmc}. All
particle decays are modeled with {\sc evtgen}~\cite{evtgen_a,evtgen_b}
using BFs either taken from the Particle Data
Group~(PDG)~\cite{pdg2022}, when available, or otherwise estimated
with {\sc lundcharm}~\cite{lundcharm_a,lundcharm_b}. Final-state
radiation from charged final-state particles is incorporated using
{\sc photos}~\cite{photos}.

Corresponding to the total number of $\psi(3686)$ events collected in
different years, 2,712,400 signal MC events are generated with
$\psi(3686) \to \pi^0h_c$ and $J/\psi \to e^+e^-/\mu^+\mu^-$ modeled
by PARTWAVE
%~(with parameters set to 1, 0, 0, 0, 0, 0 to model a decay chain of $1^-\to 1^+~0^-$) 
and PHOTOS VLL simulations~\cite{evtgen_a,evtgen_b},
respectively. Additionally, $\pi^0 \to \gamma \gamma$ and $h_c \to
\pi^+\pi^-J/\psi$ are modeled with a phase space (PHSP) model.

	%%%%%%%%%%%%%%%%%%%%%%%%%%%%%%%%%%%%%%%%%%%%%%%%%%%%%%%%%%%%%%%%
	%          3. EVENT SELECTION
	%%%%%%%%%%%%%%%%%%%%%%%%%%%%%%%%%%%%%%%%%%%%%%%%%%%%%%%%%%%%%%%%
\section{EVENT SELECTION}
\label{sec:selection}

The final-state particles in this analysis include $\pi^+~\pi^- ~\gamma ~\gamma ~\mu^+~\mu^-$ and~$\pi^+ ~\pi^- ~\gamma ~\gamma ~e^+~e^-$.
Charged tracks detected in the MDC are required to be within a polar angle ($\theta$) range of $|\rm{cos\theta}|<0.93$, where $\theta$ is defined with respect to the $z$-axis, which is the symmetry axis of the MDC. For charged tracks, the distance of closest approach to the interaction point (IP) 
must be less than 10\,cm along the $z$-axis, $|V_{z}|$,  and less than 1\,cm in the transverse plane, $|V_{xy}|$.

 Photon candidates are identified using isolated showers in the EMC.  The deposited energy of each shower must be more than 25~MeV in the barrel region ($|\cos \theta|< 0.80$) and more than 50~MeV in the end-cap region ($0.86 <|\cos \theta|< 0.92$).  To exclude showers that originate from charged tracks, the angle subtended by the EMC shower and the position of the closest charged track at the EMC must be greater than 10 degrees as measured from the IP. 
To suppress electronic noise and showers unrelated to the event, the difference between the EMC time and the event start time is required to be within [0, 700]\,ns.

%Neutral pions are reconstructed by combining two good photons and performing a one-constraint (1C) kinematic fit with the invariant mass of $\gamma \gamma$ constrained to the $\pi^0$ mass~\cite{pdg2022}.

The number of good charged tracks is required to be four with zero net
charge. The three-momenta in the laboratory frame is used to separate
leptons and pions. Charged tracks with momentum below 1~GeV$/c$ are
assumed to be pions, while those with momentum above 1~GeV$/c$ are taken
as leptons. A pair of pions with opposite charge and a pair of
leptons with the same flavor and opposite charge are required. Muons
and electrons are separated based on their energy deposits in the
EMC. Electrons and positrons must have energy deposits greater than
1.0 GeV, while muons have energy deposits less than 0.4 GeV.

The $J/\psi$ candidates are reconstructed from $e^+e^-$ and
$\mu^+\mu^-$ pairs with invariant mass in the $J/\psi$ mass region,
defined as 3.085 $<M(l^+ l^-) <$ 3.108 GeV/$c^2$,~($l \equiv e, \mu$),
which is about three times the standard deviation obtained from the fit to
the distribution of the $l^+l^-$ invariant mass $M(l^+l^-)$ of
data.

To suppress background, a five-constraint (5C) kinematic fit is
performed, constraining the four-momentum of the final state particles
to that of the initial system and the invariant mass of the
$\gamma\gamma$ pair to the $\pi^0$ mass ~\cite{pdg2022}. The four-momenta
from the kinematic fit are used in the following analysis.

To suppress background from the decay $\psi(3686) \to \eta J/\psi,
\eta \to \pi^+\pi^-\pi^0$ and other background decays to different
final states, we optimize two selection criteria, which are the
requirements on $\chi^2_{5\rm C}$ and on the $\pi^+\pi^-\pi^0$
invariant mass, $M_{\pi^+\pi^-\pi^0}$, by maximizing a figure-of-merit
given by $\epsilon_{\rm sig}/(\alpha/2+\sqrt{B})$. Here the signal
efficiency~($\epsilon_{\rm sig}$) is determined with the signal MC
sample, the background yield ($B$) is estimated using the inclusive MC
sample, and $\alpha$ is the significance level, which is set
to be 3.0. Based on the optimization, the optimal requirements of
$\chi^2_{5\rm C} < 15$ and $|M_{\pi^+\pi^-\pi^0}-M_{\eta}| >$ 24.5
MeV/$c^2$ are applied, where $M_{\eta}$ is the $\eta$
mass~\cite{pdg2022}. Additionally, a requirement of $M(\pi^+\pi^-) >$
0.3 GeV/$c^2$ has been applied to reduce the background from the decay
$\psi(3686) \to \pi^0\pi^0 J/\psi$ with $\gamma$ converting into an
$e^+e^-$ pair in the beam pipe or inner wall of the MDC and the
$e^+e^-$ pair is misidentified as $\pi^+\pi^-$ pair.  After applying
the above selection criteria, only two background events from
$\psi(3686) \to \eta J/\psi, \eta \to \pi^+\pi^-\pi^0$ survive, which
are shown as the purple bars in Fig.~\ref{fig:fit}.

	%%%%%%%%%%%%%%%%%%%%%%%%%%%%%%%%%%%%%%%%
	%        4.     SIGNAL YIELDs
	%%%%%%%%%%%%%%%%%%%%%%%%%%%%%%%%%%%%%%%%
\section{\label{Sec:BR_determined}Signal yields}
\begin{figure}[htbp]
\begin{center}
\begin{minipage}[t]{0.9\linewidth}
\includegraphics[width=1\textwidth]{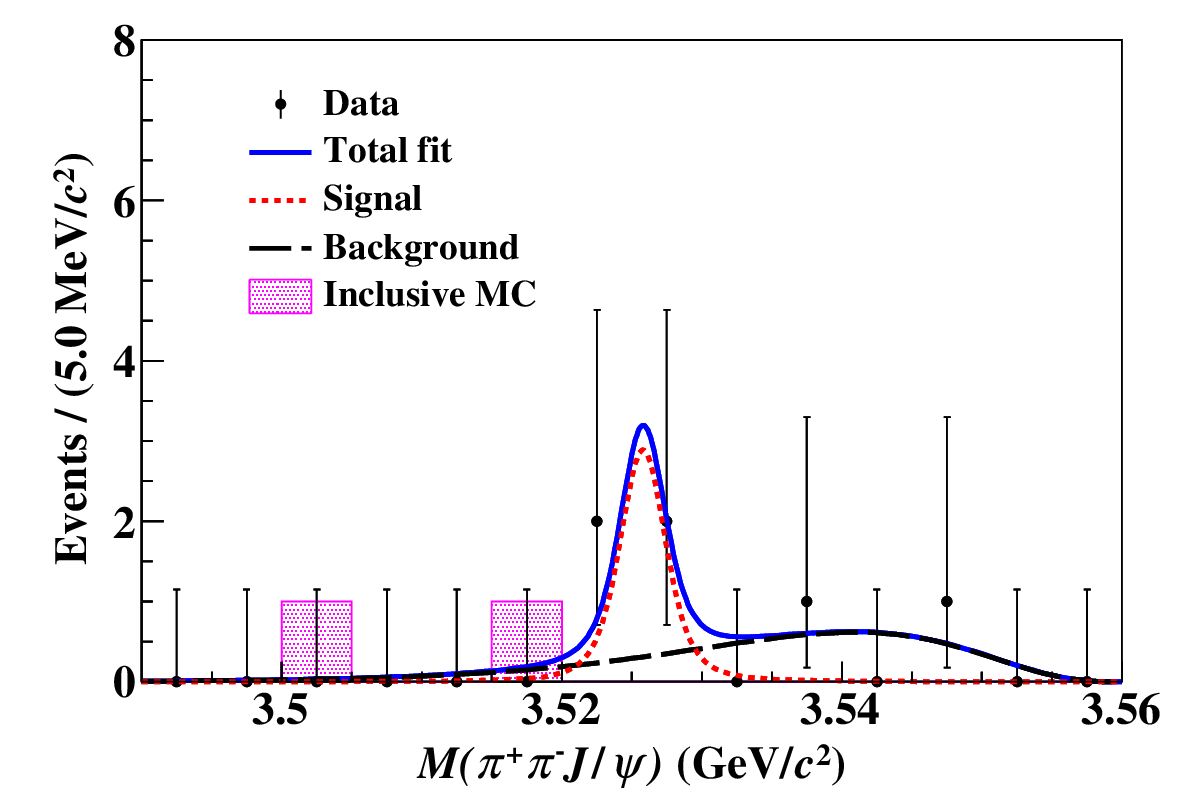}
\end{minipage}
\caption{The fit to the $M(\pi^+\pi^-J/\psi)$ distribution. The points
  with errors bar are data, the blue solid curve is the fit result,
  the black solid dashed line is the background, the red dotted curve
  is the signal and the purple bars are the remaining inclusive
  MC events.}
\label{fig:fit}
\end{center}
\end{figure}

To determine the signal yield, we perform an unbinned likelihood fit
to the $M(\pi^+\pi^-J/\psi$) distribution of the accepted candidates,
as shown in Fig.~\ref{fig:fit}. The $h_c$ signal is described using
the shape obtained from signal MC events, while the combinatorial
background shape is described by an ARGUS function~\cite{Argus}. The signal yield
determined from the fit is 
%$N^{\rm{obs}}_{\rm{sig}} = 2.7 \pm1.9$. 
$N^{\rm{obs}}_{\rm{sig}} = 2.7 \pm2.1$. 
The statistical significance of the $h_c$ signal is 1.6
$\sigma$, by comparing the log-likelihood values of the fits with and
without signal component ($\Delta \ln L$ = 1.26) and taking the change
in the number of degrees of freedom into account.

Since no significant signal of $\psi(3686) \to \pi^0h_c, h_c \to
\pi^+\pi^-J/\psi$ is observed, the upper limit on the product BFs is
calculated as

%\begin{widetext}
%\begin{equation}
%\small
%\begin{aligned}
\begin{eqnarray}
\label{equp}
[\mathcal{B}(\psi(3686) \to \pi^0h_c)\times \mathcal{B}(h_c \to
  \pi^+\pi^-J/\psi)]_{\rm upper} = \nonumber \\ 
\frac{N_{\rm sig}^{\rm up}}{N_{\psi(3686)}\times \epsilon_{sig} \times \mathcal{B}(J/\psi \to l^+l^-)\times \mathcal{B}(\pi^0 \to \gamma \gamma)}.
\end{eqnarray}
%\end{aligned}
%\end{equation}
%\end{widetext}
Here, $N_{\psi(3686)}$ is the number of $\psi(3686)$ events and
$N_{\rm sig}^{\rm up}$ is the upper limit of $h_c$ signal events
obtained from a Bayesian method~\cite{PRD012002} with a prior
density $p$, where $p=0$ for $N_{\rm sig}<0$, and $p=1$ for $N_{\rm
  sig}\geq0$. The detection efficiency of the signal mode is
$\epsilon_{\rm sig} = 3.74\%$. The BFs of the intermediate states,
$\mathcal{B}(\pi^0 \to \gamma \gamma)$ and $\mathcal{B}(J/\psi \to l^+l^-)$,
are taken from the PDG~\cite{pdg2022}. The upper limit on
$\mathcal{B}(\psi(3686) \to \pi^0h_c)\times \mathcal{B}(h_c \to
\pi^+\pi^-J/\psi)$ is determined to be \TimesBF~incorporating the
systematic uncertainties that will be addressed in
Sec.~\ref{sec:sysU}.

	%%%%%%%%%%%%%%%%%%%%%%%%%%%%%%%%%%%%%%%%
	%     6. SYSTEMATIC UNCERTAINTIES
	%%%%%%%%%%%%%%%%%%%%%%%%%%%%%%%%%%%%%%%
\section{Systematic Uncertainties}\label{sec:sysU}
The systematic uncertainty for the upper limit of the BF includes
multiplicative and additive sources.  The multiplicative systematic
uncertainties are from the number of $\psi(3686)$ events,
tracking, photon detection, kinematic fit, quoted BFs, mass windows,
and the signal model.  Details of the systematic uncertainties are discussed below:

\begin{itemize}

\item[(i)]{{Number of $\psi(3686)$ events}}: The uncertainty on the
  number of $\psi(3686)$ events, determined with inclusive
  hadronic $\psi(3686)$ decays, is 0.5 \%~\cite{CPC023001,psi2S2021}.

\item[(ii)]{{Tracking}}: The uncertainties of the tracking efficiencies
  for charged pions and leptons are estimated with a control sample
  of $\psi(3686) \to \pi^+\pi^-J/\psi, J/\psi \to
  e^+e^-~(\mu^+\mu^-)$.  The MC simulation is re-weighted in
  two-dimensional ($\cos\theta, p_{t}$) bins according to the
  efficiencies obtained from the control sample, where $\theta$ is the
  polar angle and $p_{t}$ is the transverse momentum of the charged
  pions and leptons. The corrected detection efficiency is taken as
  the nominal result. The difference in efficiencies before and after
  the correction is taken as the systematic uncertainty due to
  tracking. It is assigned to be 0.4\% for each pion and 0.1\% for
  each lepton, resulting in a total tracking uncertainty of 1.0\%.

\item[(iii)]{Photon detection}: The photon detection efficiency has
  been studied using a control sample of $e^+e^-\to \gamma
  \mu^+\mu^-$. The difference between data and MC simulation is found
  to be 0.5\% for each photon. Therefore, the total systematic
  uncertainty for two photons is assigned as 1.0\%.

\item[(iv)]{$\pi^{0}$ reconstruction}: Using a high purity control
  sample of $J/\psi \to \pi^0p\bar{p}$, the systematic uncertainty
  from $\pi^0$ reconstruction is determined to be
  1.0\%~\cite{PRL261801}.

\item[(v)]{Kinematic fit}: $\epsilon_{\rm sig}$ is the efficiency
  obtained from the signal MC sample after the helix parameter
  correction. The systematic uncertainty related to the kinematic fit
  is evaluated by comparing the efficiencies with and without this
  correction~\cite{YPG:bam}. Half of their difference, 0.7\%, is taken
  as the associated uncertainty.

\item[(vi)]{{Quoted BFs}}: The quoted BFs of $J/\psi \to e^+e^-$,
  $J/\psi \to \mu^+\mu^-$, and $\pi^0 \to \gamma \gamma$ are taken
  from the PDG~\cite{pdg2022}, with uncertainties of 0.5\%, 0.6\%, and
  0.03\%, respectively.

\item[(vii)]{Mass windows}: To evaluate the systematic uncertainty
  associated with the choice of mass windows, we perform a Barlow
  test~\cite{barlow_test} to examine the deviation in significance
  ($\zeta$) between the baseline selection and that used for the
  systematic test. The deviation in significance is defined as
\begin{equation}
    \zeta=\frac{\left|V_{\text {nominal }}-V_{\text {test }}\right|}{\sqrt{\left|\sigma_{V \text { nominal }}^2-\sigma_{V \text { test }}^2\right|}},
\end{equation}
where $V$ is $N_{\rm sig}^{\rm up}/\epsilon_{\rm sig}$;
$\sigma_{V}$ is the statistical uncertainty on $V$. The mass
windows applied on $M(\pi^+\pi^-)$ and $M(\pi^0\pi^+\pi^-)$ are
varied between $(0.270, 0.330)$~GeV/$c^2$ and $(0.225,0.305)$~GeV/$c^2$
with steps of 3.0~MeV/$c^2$ and 4.0~MeV/$c^2$, respectively. As the
$\zeta$ values are always smaller than 2, the corresponding systematic
uncertainties are ignored.

\item[(viii)]{Signal mode}: Due to the limited knowledge of the
  $M(\pi^+\pi^-)$ distribution in the decay $h_c \to
  \pi^+\pi^-J/\psi$, the signal MC sample is generated uniformly in
  phase space without considering the angular distribution in the
  nominal analysis.  To account for the potential systematic bias of
  the theoretical model, an alternative signal MC sample is generated.
  In this model, pure $P$-wave production between the two-pion system
  ($S$-wave) and $J/\psi$ is assumed, specifically $h_c \to f_0(500)
  J/\psi$, which is described by the helicity amplitude model in {\sc
    evtgen}~\cite{evtgen_a,evtgen_b}. The decay $f_0(500) \to
  \pi^+\pi^-$ is generated using a phase space model. The efficiency
  difference between these two models, 6.3\%, is assigned as the
  systematic uncertainty.
\end{itemize}

The multiplicative systematic uncertainties are summarized in Table~\ref{tab:totalsys}. The total systematic uncertainty is obtained by adding all contributions in quadrature under the assumption that they are independent.

The additive systematic uncertainties stem from the determination of the
signal yield, which depends on the fit range and signal and
background shapes. Since these additive systematic uncertainties are
correlated, we simultaneously change these three fit conditions and
choose the most conservative upper limit. The contributions
to the systematic uncertainties are discussed below.

\begin{itemize}

\item[(a)] {Fit range}: The systematic uncertainty arising from the
  fit range is evaluated by varying the fit range,
  adjusting both left and right limits by $\pm$ 10 MeV/$c^2$.

\item[(b)] {Signal shape}: The systematic uncertainty from the signal
  shape is estimated by using the MC-simulated shape convolved with a
  Gaussian function. The parameters of the Gaussian function are
  $M_{\rm Gaussian}$ = $-0.4\pm0.3$ MeV and $\sigma_{\rm Gaussian}$=
  $0.7\pm1.0$ MeV, taken from $e^+e^- \to \eta h_c, h_c \to \gamma
  \eta_c$~\cite{arXiv:2404}.

  \item[(c)] {Background shape}: The systematic uncertainty due to the
    background shape is estimated by replacing the ARGUS function~\cite{Argus} with a 0th-order polynomial
    function, or by fixing the background yield to the remaining
    inclusive MC sample.
\end{itemize}

\begin{figure}[htbp]
\begin{center}
\begin{minipage}[t]{0.9\linewidth}
\includegraphics[width=1\textwidth]{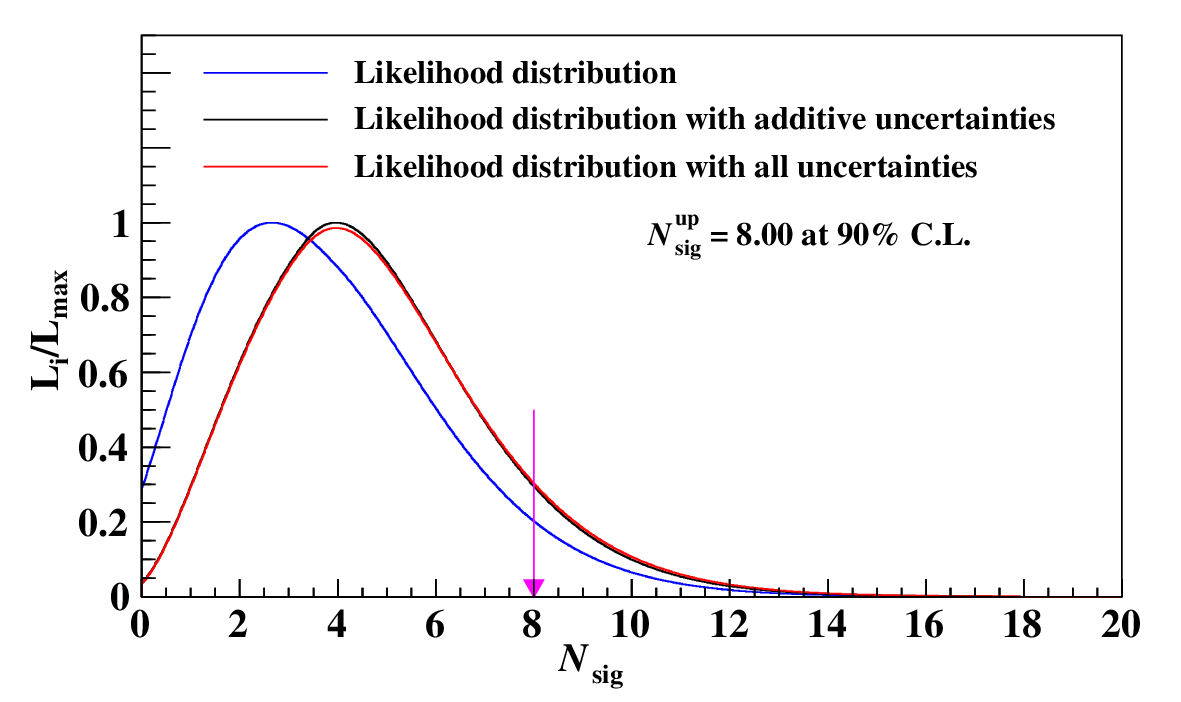}
\end{minipage}
\caption{The normalized likelihood distribution for $\psi(3686) \to
  \pi^0h_c, h_c \to \pi^+\pi^-J/\psi$. 
The result obtained without accounting for any systematic uncertainties is depicted in blue, another with additive systematic uncertainties in black, and the third considering all systematic uncertainties in red.  The arrow indicates the upper limit on the signal yield at the 90\% confidence level after accounting for all systematic uncertainties.}
\label{fig:newup}
\end{center}
\end{figure}

The multiplicative and additive systematic uncertainties are
incorporated into the calculation of the upper limit
via~\cite{Stenson:2006gwf, Liu:2015uha}
\begin{equation}
   L'(N) = \int_{\epsilon = 0}^1 {L({\textstyle{\epsilon \over {\epsilon_{\rm sig}}}}N)} \exp \left[ { - {\textstyle{{(\epsilon - \epsilon_{\rm sig})^2} \over {2\sigma _{\epsilon}^2}}}} \right]d\epsilon,
\end{equation}
where $L(N)$ is the likelihood distribution as a function of signal
yield, $N$; $\epsilon$ is the expected efficiency and $\epsilon_{\rm
  sig}$ is the nominal efficiency; $\sigma_{\epsilon}$ is its
multiplicative systematic uncertainty as summarized in
Table~\ref{tab:totalsys}. The likelihood distribution is
shown in Fig.~\ref{fig:newup}, note that the shift of the peak value is mostly due to the signal line shape uncertainty. The signal yield of the process
$\psi(3686) \to \pi^0h_c, h_c \to \pi^+\pi^-J/\psi$ at the 90\%
C.L., is taken as the upper limit ($N^{\rm up}_{\rm sig} = 8.00)$,  and used for the calculation of the
upper limit of BF.

\begin {table}[htbp]
%\begin {table*}[htbp] is middel
	\caption{Multiplicative systematic uncertainties~(\%).}
	\label{tab:totalsys}
	\begin{tabular}{lc}
	\hline
	\hline
		Source &Uncertainty \\
	\hline
		Number of $\psi(3686)$ events			&0.5\\
		Tracking  						&1.0\\
		Photon detection  					&1.0\\
		$\pi^0$ reconstruction  				&1.0\\
		Kinematic fit							&0.7\\		
		$M(\pi^+\pi^-$) mass window				&Neglected\\
		$M(\pi^+\pi^-\pi^0$) mass window  			&Neglected\\
		$\mathcal{B}(\pi^0 \to \gamma \gamma)$	&Neglected\\
		$\mathcal{B}(J/\psi \to e^+e^-)$		&0.5\\
		$\mathcal{B}(J/\psi \to \mu^+\mu^-)$	&0.6\\
		Signal model						&6.3    \\
		Total									&\Totsys    \\
        \hline
	\hline
\end{tabular}
\end{table}

	%%%%%%%%%%%%%%%%%%%%%%%%%%%%%%%%%%%%%%%%
	%        7.   SUMMARY
	%%%%%%%%%%%%%%%%%%%%%%%%%%%%%%%%%%%%%%%%
\section{Summary}
A search for the hadronic transition $h_c \to \pi^+\pi^-J/\psi$ is
performed by analyzing $(2712.4 \pm 14.3) \times 10^6
~\psi(3686)$ events collected at the BESIII experiment, no
significant signal is observed. The upper limit on the product of BFs,
that is $\mathcal{B}(\psi(3686) \to \pi^0h_c) \times \mathcal{B}(h_c
\to \pi^+\pi^-J/\psi)$, at the 90\% C.L. is determined to be \TimesBF.

Using the PDG value of $\mathcal{B}(\psi(3686)\to \pi^0
h_c)=(7.4\pm0.5)\times 10^{-4}$ and incorporating the additional
uncertainty of 6.8\%~\cite{pdg2022}, the upper limit for
$\mathcal{B}(h_c\to\pi^+\pi^-J/\psi)$ at the 90\% C.L. is estimated to
be \BF.  This result is above the predicted upper limit of
$\mathcal{B}(h_c\to\pi^+\pi^-J/\psi)$ of $1.2 \times 10^{-5}$ obtained
from the latest result of $\mathcal{B}(h_c \to
\pi^0J/\psi)$~\cite{JHEP003} as proposed by Voloshin~\cite{PRD074033}.

Neglecting the small difference between phase space for charged and
neutral $\pi \pi$ modes, we obtain $\mathcal{B}(h_c\to\pi\pi J/\psi) <
1.41\times10^{-3}$~(including charged and neutral modes) at the 90\%
C.L. by considering isospin symmetry.  The upper limit on the BF
of $h_c\to\pi\pi J/\psi$ presented in this paper is about 3 times
lower than that from the previous study~\cite{PRD052008}.  In
comparison with theoretical predictions, our result is an order of
magnitude smaller than the BF predicted by Kuang {\it et al.} (about
2\%)~\cite{PRD1210}, and closer to the prediction calculated by
Pyungwon Ko (0.05\%)~\cite{PRD1710}.
% More statistics are needed to
%improve the sensitivity and understand its underlying dynamics.
% to accurately measure the true BF of the $h_c\to\pi\pi J/\psi$ process and comprehend its underlying dynamics.

	%%%%%%%%%%%%%%%%%%%%%%%%%%%%%%%%%%%%%%%%%%%%%%%%%%%%%%%%%%%
	% ACKNOWLEDGMENTS
	%%%%%%%%%%%%%%%%%%%%%%%%%%%%%%%%%%%%%%%%%%%%%%%%%%%%%%%%%%%
	\acknowledgements
The BESIII collaboration thanks the staff of BEPCII and the IHEP computing center for their strong sup- port. This work is supported in part by National Key Research and Development Program of China under Contracts Nos. 2020YFA0406300, 2020YFA0406400; National Natural Science Foundation of China (NSFC) under Contracts Nos. 12375070, 11625523, 11635010, 11735014, 11822506, 11835012, 11935015, 11935016, 11935018, 11961141012, 12022510, 12025502, 12035009, 12035013, 12061131003; The key scientific research Projects of colleges and universities in Henan Province (21A140012); the Chinese Academy of Sciences (CAS) Large-Scale Scientific Facility Program; Joint Large-Scale Scientific Facility Funds of the NSFC and CAS under Contracts Nos. U2032108, U1732263, U1832207; CAS Key Research Program of Frontier Sciences under Contract No. QYZDJ-SSW-SLH040; the CAS Center for Excellence in Particle Physics (CCEPP); Shanghai Leading Talent Program of Eastern Talent Plan under Contract No. JLH5913002; 100 Talents Program of CAS; INPAC and Shanghai Key Laboratory for Particle Physics and Cosmology; ERC under Contract No. 758462; European Union Horizon 2020 research and innovation programme under Contract No. Marie Sklodowska-Curie grant agreement No 894790; German Research Foundation DFG under Contracts Nos. 443159800, Collaborative Research Center CRC 1044, FOR 2359, GRK 2149; Istituto Nazionale di Fisica Nucleare, Italy; Ministry of Development of Turkey under Contract No. DPT2006K-120470; National Science and Technology fund; Olle Engkvist Foundation under Contract No. 200-0605; STFC (United Kingdom); The Knut and Alice Wallenberg Foundation (Sweden) under Contract No. 2016.0157; The Royal Society, UK un- der Contracts Nos. DH140054, DH160214; The Swedish Research Council; U. S. Department of Energy under Contracts Nos. DE-FG02-05ER41374, DE-SC-0012069.

	%%%%%%%%%%%%%%%%%%%%%%%%%%%%%%%%%%%%%%%%
	%  References
	%%%%%%%%%%%%%%%%%%%%%%%%%%%%%%%%%%%%%%%%


\begin{thebibliography}{99}


  %%%%%%%%%%%%%%%%  Ref 1
  \bibitem{theory_review_a} E. Eichten, S. Godfrey, H. Mahlke, and J. L. Rosner, \href{https://journals.aps.org/rmp/abstract/10.1103/RevModPhys.80.1161}{Rev. Mod. Phys. {\bf 80}, 1161 (2008)}.


  \bibitem{hc_a} P. Rubin $et$ $al$. [CLEO Collaboration]\textit{}, \href{https://journals.aps.org/prd/abstract/10.1103/PhysRevD.72.092004}{Phys. Rev. D {\bf 72}, 092004 (2005)}.


  \bibitem{hc_b} J. L. Rosner $et$ $al$. [CLEO Collaboration]\textit{}, \href{https://journals.aps.org/prl/abstract/10.1103/PhysRevLett.95.102003}{Phys. Rev. Lett. {\bf 95}, 102003 (2005)}.

\bibitem{pdg2022} 
P. A. Zyla \textit{et. al.} (Particle Data Group), 
\href{https://academic.oup.com/ptep/article/2020/8/083C01/5891211}
{Prog. Theor. Exp. Phys. 2022, 083C01 (2022).}

\bibitem{PRD032001}
M. Andreotti \textit{et. al.} [E835 Collaboration], 
\href{https://journals.aps.org/prd/abstract/10.1103/PhysRevD.72.032001}
{Phys. Rev. D \textbf{72}, 032001 (2005).}

\bibitem{PRL102003}
J. L. Rosner \textit{et. al.} [CLEO Collaboration], 
\href{https://journals.aps.org/prl/abstract/10.1103/PhysRevLett.95.102003}
{Phys. Rev. Lett. \textbf{95}, 102003 (2005).}

\bibitem{PRL182003}
S. Dobbs \textit{et. al.} [CLEO Collaboration], 
\href{https://journals.aps.org/prl/abstract/10.1103/PhysRevLett.101.182003}
{Phys. Rev. Lett. \textbf{101}, 182003 (2008).}

\bibitem{PRD051106}
G. S. Adams \textit{et. al.} [CLEO Collaboration],
\href{https://journals.aps.org/prd/abstract/10.1103/PhysRevD.80.051106}
{Phys. Rev. D \textbf{80}, 051106 (2009).}

\bibitem{PRL132002}
M. Ablikim \textit{et. al.} [BESIII Collaboration], 
 \href{https://journals.aps.org/prl/abstract/10.1103/PhysRevLett.104.132002}
 {Phys. Rev. Lett. \textbf{104}, 132002 (2010).}
 
 \bibitem{PRD032008}
J. Y. Ge \textit{et. al.} [CLEO Collaboration],
 \href{https://journals.aps.org/prd/abstract/10.1103/PhysRevD.84.032008}
 {Phys. Rev. D \textbf{84}, 032008 (2011).}
 
 \bibitem{PRD072007}
M. Ablikim \textit{et. al.} [BESIII Collaboration], 
 \href{https://journals.aps.org/prd/abstract/10.1103/PhysRevD.106.072007}
 {Phys. Rev. D. \textbf{106}, 072007 (2022).}
 
 \bibitem{JPCS123} S. Godfrey, \href{https://iopscience.iop.org/article/10.1088/1742-6596/9/1/023} {J. Phys.: Conf. Ser. \textbf{9} 123 (2005).}
 
  \bibitem{PRD1210}
Y.~P. Kuang, S.~F. Tuan, and T.~M. Yan,
 \href{https://journals.aps.org/prd/abstract/10.1103/PhysRevD.37.1210}
{Phys. Rev. D \textbf{37}, 1210 (1988).}

 \bibitem{PRD1710}
P. Ko, 
 \href{https://journals.aps.org/prd/abstract/10.1103/PhysRevD.52.1710}
{Phys. Rev. D \textbf{52}, 1710 (1995).}

 \bibitem{PRD074033}
M. B. Voloshin, 
 \href{https://journals.aps.org/prd/abstract/10.1103/PhysRevD.86.074033}
{Phys. Rev. D \textbf{86}, 074033 (2012).}

\bibitem{CPC023001}
M. Ablikim \textit{et. al.} [BESIII Collaboration], 
~\href{https://iopscience.iop.org/article/10.1088/1674-1137/42/2/023001}
{Chin. Phys. C \textbf{52}, 023001 (2018).}

 \bibitem{PRD052008}
M. Ablikim \textit{et. al.} [BESIII Collaboration], 
 \href{https://journals.aps.org/prd/abstract/10.1103/PhysRevD.97.052008}
{Phys. Rev. D \textbf{97}, 052008 (2018).}

 \bibitem{JHEP003}
M. Ablikim \textit{et. al.} [BESIII Collaboration], 
 \href{https://link.springer.com/article/10.1007/JHEP05(2022)003}
{JHEP \textbf{05}, 003 (2022).}


\bibitem{psi2S2021}
M. Ablikim \textit{et. al.} [BESIII Collaboration], 
~\href{https://iopscience.iop.org/article/10.1088/1674-1137/ad595b}
{Chin. Phys. C \textbf{48}, 093001 (2024).}

  %%%%%%%%%%%%%%%%   Ref 15
  \bibitem{Ablikim:2009aa} M. Ablikim $et$ $al$. [BESIII Collaboration], \href{https://doi.org/10.1016/j.nima.2009.12.050}{Nucl. Instrum. Meth. A {\bf 614}, 345 (2010)}.

  \bibitem{CXYu_bes3} C.~H.~Yu {\it et al.}, Proceedings of IPAC2016, Busan, Korea (JACoW, Busan, 2016), \url{https://accelconf.web.cern.ch/ipac2016/}.


  \bibitem{Ablikim:2019hff} M.~Ablikim {\it et al.} [BESIII Collaboration], \href{https://iopscience.iop.org/article/10.1088/1674-1137/44/4/040001}{Chin. Phys. C {\bf 44}, 040001 (2020)}.

  \bibitem{EcmsMea}
  J.~Lu, Y.~Xiao, and X.~Ji, \href{https://doi.org/10.1007/s41605-020-00188-8}{Radiat. Detect. Technol. Methods {\bf 4}, 337 C344 (2020)}.
  %https://doi.org/10.1007/s41605-020-00188-8

  \bibitem{EventFilter}
  J.~W.~Zhang {\it et al.}, \href{https://doi.org/10.1007/s41605-022-00331-7}{Radiat. Detect. Technol. Methods {\bf 6}, 289 C293 (2022)}.
  %https://doi.org/10.1007/s41605-022-00331-7


   %%%%%%%%%%%%%%%%  Ref 20
  \bibitem{tof_a} X. Li \textit{et al.}, \href{https://link.springer.com/article/10.1007\%2Fs41605-017-0014-2}{Radiat. Detect. Technol. Meth. {\bf 1} 13 (2017)}.


  \bibitem{tof_b} Y. X. Guo \textit{et al.}, \href{https://link.springer.com/article/10.1007\%2Fs41605-017-0012-4}{Radiat. Detect. Technol. Meth. {\bf 1} 15 (2017)}.

  \bibitem{tof_c} P. Cao \textit{et al.}, \href{https://www.sciencedirect.com/science/article/pii/S0168900219314068?via\%3Dihub}{Nucl. Instrum. Meth. A {\bf 953}, 163053 (2020)}.

   \bibitem{geant4} S.~Agostinelli \textit{et al.} [GEANT4 Collaboration], \href{https://doi.org/10.1016/S0168-9002(03)01368-8}{Nucl. Instrum. Meth. A {\bf506}, 250 (2003).}

%   \bibitem{detvis} K.~X.~Huang {\it et al.}, \href{https://doi.org/10.1007/s41365-022-01133-8}{Nucl.\ Sci.\ Tech. {\bf 33}, 142 (2022)}.

  %%%%%%%%%%%%%%%%  Ref 25
   \bibitem{kkmc_a} S. Jadach, B. F. L. Ward, and Z. Was, \href{https://www.sciencedirect.com/science/article/pii/S0010465500000485?via\%3Dihub}{Comput. Phys. Commun. {\bf 130}, 260 (2000)}.


    \bibitem{kkmc_b} S. Jadach, B. F. L. Ward, and Z. Was, \href{https://journals.aps.org/prd/abstract/10.1103/PhysRevD.63.113009}{Phys. Rev. D {\bf 63}, 113009 (2001)}.


    \bibitem{evtgen_a} R. G. Ping, \href{https://iopscience.iop.org/article/10.1088/1674-1137/32/8/001}{Chin. Phys. C {\bf 32}, 599 (2008)}.


   \bibitem{evtgen_b} D. J. Lange, \href{https://www.sciencedirect.com/science/article/pii/S0168900201000894?via\%3Dihub}{Nucl. Instrum. Meth. A {\bf 462}, 152 (2001)}.

%\bibitem{pdg2022} 
%P. A. Zyla \textit{et. al.} (Particle Data Group), 
%\href{https://academic.oup.com/ptep/article/2020/8/083C01/5891211}
%{\blue{Prog. Theor. Exp. Phys. 2022, 083C01 (2022).}}

    \bibitem{lundcharm_a} J. C. Chen, G. S. Huang, X. R. Qi, D. H. Zhang, and Y. S. Zhu, \href{https://journals.aps.org/prd/abstract/10.1103/PhysRevD.62.034003}{Phys. Rev. D {\bf 62}, 034003 (2000)}.

  %%%%%%%%%%%%%%%%  Ref 30
  \bibitem{lundcharm_b} R. L. Yang, R. G. Ping, and H. Chen, \href{https://iopscience.iop.org/article/10.1088/0256-307X/31/6/061301}{Chin. Phys. Lett. {\bf31}, 061301 (2014)}.

  \bibitem{photos} E. Richter-Was, \href{https://www.sciencedirect.com/science/article/pii/037026939390062M?via\%3Dihub} {Phys. Lett. B {\bf 303}, 163 (1993). }

 \bibitem{Argus}
ARGUS Collaboration, 
 \href{https://journals.aps.org/prl/abstract/10.1103/PhysRevLett.105.261801}
{Phys. Lett. B \textbf{241}, 278 (1990).}


\bibitem{PRD012002} 
J. Conrad, O. Botner, A. Hallgren and C.~Perez de los Heros,
\href{https://journals.aps.org/prd/abstract/10.1103/PhysRevD.67.012002}
{Phys. Rev. D \textbf{67}, 012002 (2003).}

 \bibitem{PRL261801}
M. Ablikim \textit{et. al.} [BESIII Collaboration], 
 \href{https://journals.aps.org/prl/abstract/10.1103/PhysRevLett.105.261801}
{Phys. Rev. Lett \textbf{105}, 261801 (2010).}

 %\bibitem{zhouxy_topoAna} X. Y. Zhou, S. X. Du, G. Li and C. P. Shen, \href{https://doi.org/10.1016/j.cpc.2020.107540}{Comput. Phys. Commun. {\bf 258}, 107540 (2021)}.


  %\bibitem{Argus} H. Albrecht \textit{et al}. \href{https://www.sciencedirect.com/science/article/pii/0370269394013020?via\%3Dihub}{Phys. Lett. B {\bf 340}, 217 (1994)}.


  %\bibitem{PhysRevD.57.3873} Gary J. Feldman and Robert D. Cousins, \href{https://journals.aps.org/prd/abstract/10.1103/PhysRevD.57.3873}{Phys. Rev. D {\bf57}, 3873 (1998)}.


  %%%%%%%%%%%%%%%%  Ref 35
 %\bibitem{error_ppb2pi} W. L. Yuan, X.-C. Ai, X.-B. Ji, S.-J. Chen, Y. Zhang, L.-H. Wu, L.-L. Wang, and Y. Yuan, \href{https://iopscience.iop.org/article/10.1088/1674-1137/40/2/026201}{Chin. Phys. C {\bf40}, 026201 (2016)}.

%\bibitem{PRD112005} 
%M. Ablikim \textit{et. al.} (BESIII Collaboration), 
%\href{https://journals.aps.org/prd/abstract/10.1103/PhysRevD.83.112005}
%{\blue{Phys. Rev. D \textbf{83}, 112005 (2011).}}

  \bibitem{YPG:bam} M. Ablikim $et$ $al$. [BESIII Collaboration],\textit{} \href{https://journals.aps.org/prd/abstract/10.1103/PhysRevD.87.012002}{Phys. Rev. D {\bf87}, 012002 (2013)}.


  \bibitem{barlow_test} O. Behnke, K. Kroninger, G. Schott, and T. H. Schorner-Sadenius, \href{https://www.wiley.com/en-cn/Data+Analysis+in+High+Energy+Physics\%3A+A+Practical+Guide+to+Statistical+Methods-p-9783527410583}{(Wiley-VCH, Berlin, Germany, 2013)}.

  %%%%%%%%%%%%%%%%  Ref 39

  \bibitem{arXiv:2404} M. Ablikim \textit{et. al.} [BESIII Collaboration],  \href{https://arxiv.org/abs/2404.06718}{arXiv: hep-ex/2404.06718}.
  
  \bibitem{Stenson:2006gwf} K. Stenson, \href{https://arXiv.org/abs/physics/0605236}{arXiv: physics/0605236}.


  \bibitem{Liu:2015uha} X. X. Liu, X. R. Lyu, and Y. S. Zhu, \href{https://doi.org/10.1088/1674-1137/39/10/103001}{Chin. Phys. C {\bf 39}, 113001 (2015)}.




\end{thebibliography}
\end{document}